\pdfoutput=1

\documentclass[aps,prd,twocolumn,groupaddress,tightenlines,nofootinbib,showpacs]{revtex4-1}

\usepackage{amsfonts,amsmath,amsthm,amssymb,txfonts}
\usepackage{mathrsfs}
\usepackage{bm}
\usepackage{color}
\usepackage{epsfig}
\usepackage{mathrsfs}

\usepackage{delarray}
\usepackage{dcolumn}

\usepackage[usenames,dvipsnames]{xcolor}
\definecolor{orange}{cmyk}{0,0.5,1,0}
\definecolor{rossoCP3}{cmyk}{0,.88,.77,.40}
\definecolor{graa}{rgb}{0.8,0.8,0.8}
\definecolor{blaa}{rgb}{0.2,0.2,0.6}

\newcommand{\met} {\not\!\! E_T}

\newcommand{\beq}{\begin{equation}}
\newcommand{\eeq}{\end{equation}}
\newcommand{\bea}{\begin{flushleft} \begin{eqnarray}}
\newcommand{\eea}{\end{eqnarray}\end{flushleft}}

\newcommand{\postscript}[2]{\setlength{\epsfxsize}{#2\hsize}
   \centerline{\epsfbox{#1}}}
\newcommand{\comment}[1]{}

\newcommand{\ci}[1]{}

\newcommand{\ba}{\begin{eqnarray}}
\newcommand{\ea}{\end{eqnarray}}
\newcommand{\be}{\begin{equation}}
\newcommand{\ee}{\end{equation}}
\newcommand{\bay}[1]{\left(\begin{array}{#1}}
\newcommand{\eay}{\end{array}\right)}

\def\met{\mbox{${\hbox{$E$\kern-0.6em\lower-.1ex\hbox{/}}}_T$}} 

\newcommand{\beqa}{\begin{eqnarray}}
\newcommand{\eeqa}{\end{eqnarray}}

\begin{document}

\title{\color{rossoCP3}{Neutrino lighthouse powered by Sagittarius
    A$^{\bm{\!*}}$ disk dynamo}}
				
\author{Luis A. Anchordoqui}
\affiliation{Department of Physics and Astronomy, Lehman College, City University of
  New York, NY 10468, USA\\
Department of Physics, Graduate Center, City University
  of New York, 365 Fifth Avenue, NY 10016, USA\\
Department of Astrophysics, American Museum of Natural History, Central Park West  79 St., NY 10024, USA}


\begin{abstract}\vskip 2mm
  \noindent We show that the subset of high energy neutrino events
  detected by IceCube which correlate with the Galactic center (within
  uncertainties of their reconstructed arrival directions) could
  originate in the collisions of protons accelerated by the
  Sagittarius (Sgr) A $^{\!*}$ disk dynamo. Under very  reasonable
  assumptions on source parameters we demonstrate that the
  supermassive black hole at the center of the Galaxy could launch
  protons and nuclei with multi PeV energies. Acceleration of these
  particles in a period of seconds up to Lorentz factors of $\sim
  10^{7.5}$ is possible by means of the Blandford-Znajek mechanism, which
  wires the spinning magnetosphere of Sgr A$^*$ as a Faraday unipolar
  inductor. During the acceleration process the $\sim {\rm PeV}$
  progenitors of $\sim 50~{\rm TeV}$ neutrinos radiate curvature
  photons in the keV energy range. We show that IceCube neutrino
  astronomy with photon tagging on the Chandra X-ray Observatory could
  provide a valuable probe for the Blandford-Znajek acceleration
  mechanism. We also argue that EeV neutrinos, which may be produced
  in a similar fashion during the merging of binary black holes, could
  become the smoking gun for particle acceleration in a one-shot boost.

\end{abstract}
 

\maketitle

The IceCube high-energy starting event (HESE) search resulted in the
discovery of astrophysical neutrinos~\cite{Aartsen:2013bka}.  In four years of data taking,
54 events with neutrino vertex contained in the detector volume have been
observed~\cite{Aartsen:2014gkd}. The HESE sample rejects a purely atmospheric explanation at
more than $5.7\sigma$. The data are consistent with
expectations for equal fluxes of all three neutrino
flavors~\cite{Aartsen:2015ivb}.  While the flux above 200~TeV can be
accommodated by a power law with a spectral index $\gamma = 2.07 \pm
0.13$~\cite{TeVPA}, lowering the threshold revealed an excess of
events in the $30 - 200$~TeV energy range~\cite{Aartsen:2014muf},
raising the possibility that the cosmic neutrino spectrum does not
follow a single power law, and/or may be contaminated by additional
charm background~\cite{Halzen:2016pwl}.

Quite recently, the IceCube Collaboration reported a combined analysis
based on six different searches for astrophysical neutrinos (that
included the HESE data sample)~\cite{Aartsen:2015knd}. Assuming the
neutrino flux to be isotropic and to consist of equal flavors at
Earth, the all flavor spectrum with neutrino energies $25~{\rm TeV}
\leq \varepsilon_\nu \leq 2.8~{\rm PeV}$ is well described by an unbroken power
law with best-fit spectral index $-2.50 \pm 0.09$ and a flux at
100~TeV of $6.7^{+1.1}_{-1.2} \times 10^{-18}~({\rm GeV\, s \, sr \,
  cm}^2)^{-1}$. Splitting the data into two sets, one from the
northern sky and one from the souther sky, allows for a satisfactory
power law fit with a different spectral index for each hemisphere.
The best-fit spectral index in the northern sky is $\gamma_N =
2.0^{+0.3}_{-0.4}$, whereas in the southern sky it is $\gamma_S = 2.56
\pm 0.12$. The discrepancy with respect to a single power law
corresponds to $1.1 \sigma$~\cite{Aartsen:2015knd} and may indicate that the
neutrino flux is anisotropic~\cite{Neronov:2015osa}. 

The higher statistics data seem to reinforce earlier claims that a
significant part of the flux contributing to the HESE sample
originates in the Galaxy~\cite{Fox:2013oza}. In particular, it has
been noted that the largest HESE concentration is at or near the
Galactic center, within uncertainties of their reconstructed arrival
directions~\cite{Razzaque:2013uoa,Bai:2014kba}. The excess contains 7
shower-like events, including a PeV event ($\# 14$) and two sub-PeV
events ($\#2$ and $\#25$) with arrival directions consistent with a
Galactic center origin within their $1\sigma$ uncertainty.
Interestingly, 2 of the 7 events occurred within one day of each other
and there is a 1.6\% probability that this would occur for a random
distribution in time~\cite{Bai:2014kba}.  More compelling, event
$\#25$ has a time very close to (around three hours after) the
brightest X-ray flare of Sagittarius (Sgr) A$^*$ observed by the
Chandra X-ray Observatory~\cite{Nowak:2012ry}, with a $p$-value of
0.9~\cite{Bai:2014kba}.  In this paper we show that this intriguing
correlation could provide the first experimental evidence favoring the
Blandford-Znajek mechanism, which wires the spinning magnetosphere of
Sgr A$^*$ as a Faraday unipolar inductor~\cite{Blandford:1977ds}. We
first investigate whether the supermassive black hole at the center of
the Galaxy could be the source of multi-PeV cosmic rays that would
produce high-energy gamma rays and neutrinos upon interaction with the
ambient gas. After that we further examine radiative processes which
lead to energy losses during the acceleration.  In particular, we
study radiation of curvature photons on the black hole
magnetosphere. Under reasonable assumptions on source parameters we
show that protons accelerated by the spinning black hole dynamo could
explain the IceCube-Chandra connection advanced in~\cite{Bai:2014kba}.

To develop some sense for the orders of magnitude involved, recall
that the maximum luminosity of an accreting black hole is given by the
Eddington limit,
\begin{equation}
L_{\rm Edd} \sim 1.3 \times 10^{44} M_6~{\rm erg} \, {\rm s}^{-1} \,,
\end{equation}
where $M = 10^6 M_6 M_\odot$ is the black hole mass and $M_\odot$ is
the solar mass. At any higher luminosity, ordinary matter is more
likely to be driven off by radiation pressure than to accrete. The
critical accretion rate required to sustain the Eddington luminosity,
assuming a typical $\eta = 0.1$ efficiency of conversion of mass to
radiant energy, is found to be
\begin{equation}
\dot{M}_{\rm Edd} = \frac{L_{\rm Edd}}{\eta c^2} \simeq 2.2 \times
10^{-2} \ M_6 \ M_\odot~{\rm yr}^{-1} \, .
\end{equation}
 The accreting plasma is assumed to support an axisymmetric magnetic field
configuration due to the generation of currents. The characteristic
strength of the $\vec B$-field can be obtained assuming pressure
equilibrium between the magnetic field and the in-falling matter,
whereby~\cite{Boldt:1999ge}
\begin{equation}
B_p \sim 6 \times 10^5 \ M_6^{-1/2} \ \left(\frac{\dot M}{\dot M_{\rm Edd}}
\right)^{1/2}~{\rm G}.
\label{Bp}
\end{equation}
Accretion from a thin disk would spin a black hole up to its near
critical angular momentum 
\begin{equation}
J = a \, J_{\rm max} = a \, \frac{GM^2}{c} \,,
\end{equation}
where $a$ is the dimensionless spin parameter. Now a point worth noting at this juncture is that for a Schwarzschild
black hole, $a = 0$, whereas for a maximally spinning (extreme Kerr)
black hole, $a =1$. Actually, the increase of the spin parameter would
stop at $a \approx 0.998$ due to the fact that photons emitted from
the disk on retrograde paths are more likely to be captured by the
hole than prograde ones, hence de-spinning the hole~\cite{Thorne:1974ve}.

For static black holes, the scale characterizing
the event horizon $r_h$ coincides with the Schwarzschild
radius
\begin{equation}
r_S = \frac{2GM}{c^2} \simeq 3 \times 10^{11} \, M_6~{\rm cm} \, .
\end{equation}
For spinning black holes, the (spherical) event horizon surface has a radius
\begin{equation}
r_h = r_g \left(1 + \sqrt{1-a^2} \right) \, ,
\end{equation}
where $r_g = r_S/2$ is the gravitational radius.  Just {\it outside}
the event horizon lies the ergoregion defined by
\begin{equation}
r_h < r < r_g \left(1 + \sqrt{1-a^2 \cos^2 \theta} \right) \,,
\end{equation}
where $\theta$ is the angle to the polar
axis. The outer boundary of the ergoregion is the ergosphere, which is ellipsoidal in shape
and meets the horizon at the poles $\theta = 0, \pi$. Inside the ergosphere, spacetime
is dragged along in the direction of the black hole rotation (frame
dragging), so that no static observer can exist and a particles must 
co-rotate with the hole. The angular velocity $\Omega$ of a
black hole is defined as the angular velocity of the dragging of
inertial frames at the horizon, and is found to be
\begin{equation}
\Omega = a \left(\frac{c}{2 r_h} \right) \, .
\end{equation}

The supermassive spinning black hole endowed with the external poloidal field $B_p$
induces an electric field of magnitude $E⃗ \sim \Omega
r_h B_p/c$.  This has associated a permanent voltage drop
across the horizon of magnitude 
\begin{equation}
\Phi  \sim  E r_h = \frac{a r_h B_p}{2}
 \sim  2 \times 10^{17} a \left(1 + \sqrt{1-a^2}\right) M_6 B_4~{\rm V},
\end{equation} 
where $B_p = 10^4 B_4~{\rm G}$. In analogy with an electric circuit,
the black hole can be thought as a disk dynamo with non-zero
resistance, so that power can be extracted by currents flowing between
its equator and poles~\cite{Blandford:1977ds}.
If a cosmic ray nucleus of charge $Z$ can fully tap this potential, acceleration up to
$\varepsilon = Ze \Phi \sim 10^{2.5} Z M_6 B_4~{\rm PeV}$ may
become possible, where we have taken $a \sim 1$.  However, the charge density in the vicinity of accreting black holes
could be so high that a significant fraction of this potential would be
screened and so no longer available for particle
acceleration. Therefore it seems more appropriate to define an
effective potential where the available length scale,  the gap height $\zeta$, is
explicitly taken into account~\cite{Rieger:2011ch}
\begin{equation}
\Phi_{\rm eff} \sim \left(\frac{\zeta}{r_h}\right)^2 \Phi =
\frac{\Omega}{c} \left(\frac{\zeta}{r_h}\right)^2 r_h^2 B_p \, .
\end{equation}
In the absence of energy losses the maximum Lorentz factor is given by
\begin{equation}
\gamma_{\rm max}^{\rm acc} \sim 10^{8.5} \frac{Z M_6 B_4}{A}
\left(\frac{\zeta}{r_h} \right)^2 \, ,
\label{gamma-cr}
\end{equation}
 where $A$ is the nucleus mass number.
Accordingly, the characteristic rate of energy
gain is found to be
\begin{equation}
\left. \frac{d\varepsilon}{dt} \right|_{\rm acc} = \frac{Z e \, c \, \Phi_{\rm eff}}{\zeta}  \, .
\end{equation}
Within the potential drop, the cosmic rays  follow the curved magnetic
field lines and so emit curvature-radiation photons. The energy loss
rate or total power radiated away by a single cosmic ray is~\cite{Ochelkov} 
\begin{equation}
\left. \frac{d \varepsilon}{dt} \right|_{\rm loss} = \frac{2}{3} \frac{Z^2 e^2 c}{r_c^2}
\gamma^4 \,,
\end{equation}
where $r_c$ is the curvature radius of the magnetic field lines, and
$\gamma$ is the Lorentz factor of the radiating particles. 
Acceleration gains are balanced by radiative losses. In the
absence of other damping mechanisms,  the radiation reaction limit
is~\cite{Levinson:2000nx} 
\begin{equation}
\gamma_{\rm max}^{\rm rad} \sim 10^9 \left(\frac{a B_4 }{Z} \right)^{1/4} M_6^{1/2}
\left(\frac{r_c}{r_h} \right)^{1/2}
\left(\frac{\zeta}{r_h}\right)^{1/4} \, .
\end{equation}
All in all, direct electric field acceleration on the black hole magnetosphere would allow Lorentz factors up to
\begin{equation}
\gamma_{\rm max} = {\rm min} \, \left\{\gamma_{\rm max}^{\rm acc}, \
\gamma_{\rm max}^{\rm rad} \right\}\, .
\end{equation}
The resulting cosmic ray spectrum would be sharply peaked at nearly the
maximum energy, like spectra of high-energy particles
produced in linear accelerators. This particular footprint may help
distinguishing a Blandford-Znajek origin from conventional Fermi shock
acceleration, which typically gives power law spectra $\propto \varepsilon^{-2}$~\cite{Neronov:2007mh}.
The curvature radiation spectrum emitted by mono-energetic particles has a peak at
\begin{equation}
\varepsilon_{\gamma, {\rm max}} = \frac{3}{2} \, \hbar c \, \frac{\gamma_{\rm max}^3}{r_c}
\, .
\label{gamma-gamma}
\end{equation}
For $\gamma \ll \gamma_{\rm max}$, the curvature spectrum radiated by
single particle follows a power law, $\varepsilon_\gamma^{1/3}$, as in
synchrotron emission, and then decreases exponentially for $\gamma \gg
\gamma_{\rm max}$. 

\begin{figure*}[tpb]
\begin{minipage}[t]{0.49\textwidth}
\postscript{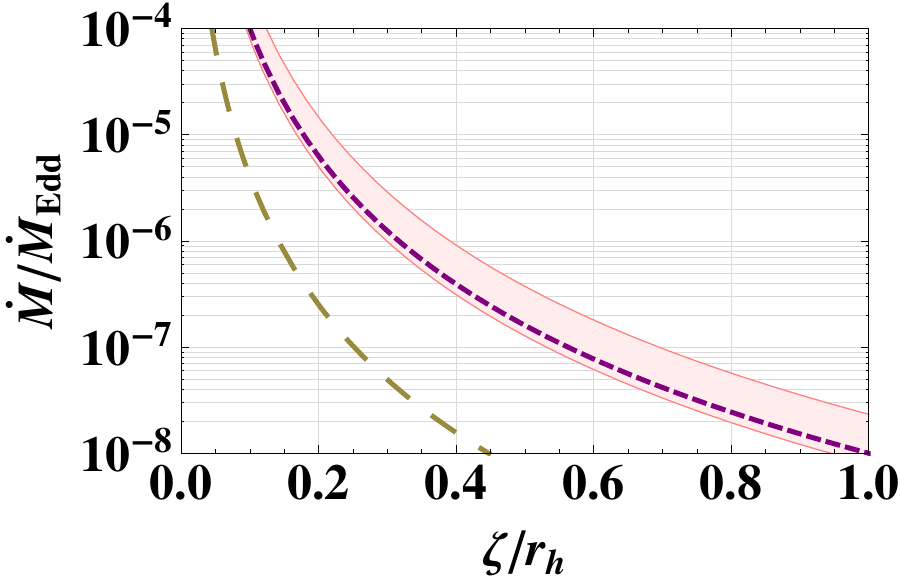}{0.9}
\end{minipage}
\begin{minipage}[t]{0.49\textwidth}
\postscript{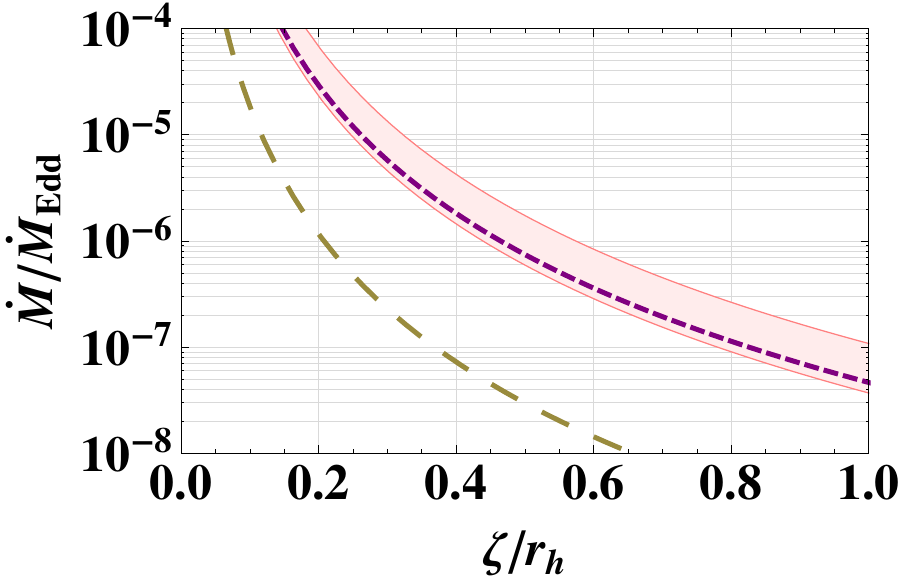}{0.9}
\end{minipage}
\caption{The pink band indicates the region of the parameter space
  that can explain the Chandra flare in the keV energy range from
  curvature radiation of protons (left) and iron nuclei (right). The
  dashed lines (long $\varepsilon_\nu/\varepsilon = 10^{-1.3}$ and
  short $\varepsilon_\nu/\varepsilon = 10^{-2}$) are contours of
  constant Lorentz factors of protons (left) and nuclei (right), which
  could produce a $\sim 35~{\rm TeV}$ neutrino after scattering on the
  ambient gas. We have taken $r_c \sim r_g$. We avoid reference to specific accretion models and
  present results for the generous range $10^{-9}~{\rm yr}^{-1} < \dot M/M_\odot <
  10^{-4}~{\rm yr}^{-1}$, which  can accommodate the X-ray luminosity
  observed by Chandra: $L_{\rm X} \sim 5 \times 10^{35}~{\rm erg \, s^{-1}}$.}
\label{fig}
\end{figure*}

At this point a reality check is in order. Sgr A$^*$ has a mass of $M
\simeq 4.31 \times 10^6 M_\odot$~\cite{Gillessen:2008qv} and the
accretion rate varies in the range $2 \times 10^{-9}~{\rm yr}^{-1}
\alt \dot M/M_\odot \alt 2 \times 10^{-7}~{\rm
  yr}^{-1}$~\cite{Quataert:2000xya} (see
however~\cite{Narayan:1997ku}). Altogether this yields a peak on the
photon spectrum at
\begin{equation}
\varepsilon_{\gamma, {\rm max}} = 2.8 \times 10^{12} \, \left(\frac{\dot M}{\dot M_{\rm Edd}}\right)^{3/2} \left(\frac{Z}{A} \right)^3 \,
\left(\frac{\zeta}{r_h}\right)^6~{\rm keV} \, .
\end{equation}
The highest peak flux and fluency flare ever observed from Sgr A$^*$
lasted for about 5.6~ks  and had a
decidedly asymmetric profile with a faster decline than
rise~\cite{Nowak:2012ry}.  The total $2 \alt \varepsilon_\gamma/{\rm
  keV}\alt 10$ emission of the event was approximately $10^{39}~{\rm
  erg}$.  Partially ordered magnetic fields with intrahour
variability~\cite{Johnson:2015iwg}~could partially accommodate the
duration of the flare. The $\sim \varepsilon_\gamma^{-2}$ spectrum
observed by Chandra over less than a decade of
energy~\cite{Nowak:2012ry} can be accommodated by tuning appropriate
source parameters, but it is less naturally obtained than with Fermi
mechanism~\cite{Fermi:1949ee}. The IceCube event $\#25$ correlating
with the Sgr A$^*$ flare has a shower topology, and therefore the
deposited electromagnetic equivalent energy in the detector,
$33.5^{+4.9}_{-5.0}$, is roughly equal to the neutrino energy
$\varepsilon_\nu \approx 35~{\rm TeV}$~\cite{Aartsen:2014gkd}.
Backgrounds to astrophysical neutrinos originate in cosmic ray air
showers. Muons produced in these showers, mainly from decay of charged
pions and kaons, enter the detector from above. The total 
atmospheric muon background in four years of data is found to be $12.6
\pm 5.1$~\cite{Aartsen:2014gkd}.  The background from atmospheric
neutrinos reported by the IceCube Collaboration is
$9.0^{+8.0}_{-2.2}$~\cite{Aartsen:2014gkd}. Note that neutrino
background events are overwhelmingly $\nu_\mu$~\cite{Aartsen:2012uu},
which produce track topologies. For this reason, the chance that the
HESE event $\#25$ is atmospheric in origin is negligible. Sgr A$^*$ is
radiatively inefficient, so the photon density will be low in its
environment.  The protons and/or nuclei accelerated by the radiatively
inefficient accretion flow (RIAF) escape the engine and interact with
the dense molecular gas surrounding Sgr A$^{\!*}$. This gas
concentration is known as the central molecular zone (CMZ), with a
size of $R_{\rm CMZ} \sim100~{\rm pc}$ and a mass of $M_{\rm CMZ} \sim
10^7 M_\odot$~\cite{Morris:1996th}. Neutrino production proceeds
dominantly through $pp$ collisions within the
CMZ~\cite{Fujita:2015xva}. Then, the neutrino energy is expected to
scale with that of the parent protons according to $10^{-2} \alt
\varepsilon_\nu/\varepsilon \alt
10^{-1.3}$~\cite{Kelner:2006tc}. Substituting these figures in
(\ref{gamma-cr}) and (\ref{gamma-gamma}) we can study the required
conditions to associate the IceCube neutrino event with the flare
observed by Chandra. Our results are encapsulated in Fig.~\ref{fig},
where we show that the possible connection is valid for a large range
of $\dot M$ and $\zeta$. Note that for $\zeta \sim r_g$, $Z= A = 1$,
and $\dot M \sim 5 \times 10^{-8} M_\odot~{\rm yr}$, we find
$\gamma_{\rm max} \sim 10^{7.5}$. Interaction of these energetic
protons within the CMZ could lead to PeV neutrinos.

Thus far we have shown (under reasonable assumptions) that PeV and
sub-PeV neutrinos could be produced in the vicinity of the Galactic
center through $pp$ collisions and that the parent protons could also
radiate X-rays during the acceleration process. To explain the 2:38:29
hours delay of IceCube detection of event \#25 with respect to the
Chandra giant flare we require further assumptions.  In particular,
the $pp$ collisions should take place before the ultra-relativistic
protons undergo diffusion in the CMZ. The diffusion properties depend
strongly on the magnetic field strength, which is quite
uncertain~\cite{YusefZadeh:2012nh,Yoast-Hull:2014cra}. Early estimates
inferred a large-scale milligauss magnetic field permeating throughout
the Galactic center based on the apparent resistance of nonthermal
filaments to distortion by molecular clouds~\cite{YusefZadeh:1984}.
More recent estimates are somewhat lower: a strength of $\sim
6~\mu{\rm G}$ was inferred from radio emission distributed over the
inner $6^\circ \times 2^\circ$~\cite{LaRosa:2005ai}, while a strength
$\agt 50~\mu {\rm G}$ was deduced from nonthermal radio emission in
the inner $3^\circ \times 2^\circ$ of the
Galaxy~\cite{Crocker:2010xc}.  Herein we
follow~\cite{YusefZadeh:2012nh} and adopt a fiducial value on the
extreme lower side, $B_{\rm CMZ} \sim 10~\mu {\rm G}$.  For
$\varepsilon \sim 3.5~{\rm PeV}$, this leads to a proton Larmor radius
\begin{equation}
r_L \simeq 1.1 \frac{(\varepsilon/{\rm PeV})}{(B_{\rm CMZ}/\mu{\rm
    G})}~{\rm pc} \sim 0.4~{\rm pc} \, .
\end{equation}
The average gas density of the CMZ is estimated to be $10^{4}~{\rm
  cm}^{-3}$~\cite{Jones:2013kra}. In CMZ clouds (which have a
line-of-sight length of 1~pc) the gas density may increase up to about
$10^6~{\rm cm}^{-3}$~\cite{Immer:2016}. Accordingly, we assume
$pp$ interactions occur within the CMZ innermost region, where the gas
density could be up to one~\cite{Kistler:2015oae} to
two~\cite{Aharonian:2004jr} orders of magnitude higher. We adopt the
$pp$ cooling time advocated in~\cite{Aharonian:2004jr}
\begin{equation}
t_{pp} = \frac{1}{\sigma_{pp} \; n \; c} \sim 7 \times 10^{6}~{\rm s} \, ,
\end{equation}
where $n \sim 10^{8}~{\rm cm}^{-3}$ is the number density of the accretion
plasma and $\sigma_{pp} \sim 50~{\rm mb}$ is  the  inelastic cross
section. Note that for this particular choice of parameters the proton mean free
path is $ct_{pp} < r_L$, and so the particles could interact before they
diffuse in the CMZ. Needless to say, one may argue that to
accommodate the data we have made very speculative considerations in
regards to $n$ and $B_{\rm CMZ}$. Whichever point of view one may find
more convincing, it seems most conservative at this point to depend on
experiment (if possible) to resolve the issue.

Electrons are also accelerated in the inner parts of Sgr A$^*$. For
such extremely low luminosity RIAF, the energy losses are dominated by
synchrotron and curvature emission (inverse Compton scattering
dominates in relatively high luminosity
RIAFs)~~\cite{Ptitsyna:2015nta}. The radiative energy loss rate of
electrons is much higher than that of protons, and so accelerated
electrons provide more economic ways for production of high-energy
gamma rays, with spectra extending into the 100~GeV energy
range~\cite{Aharonian:2004jr}. It seems then reasonable to associate
the Chandra flare with proton curvature radiation, rather than with
electron processes. Future correlation studies of new IceCube data and Chandra
observations~\cite{Ponti:2015tva} would provide a definitive probe of this model.

Now, if the Blandford-Znajek mechanism operates at the center of our
Galaxy it is very likely that it is also at play in most active
galaxies. Hence, it is tempting to speculate that the putative
association of high energy neutrinos with blazar
flares~\cite{Halzen:2005pz} could also be indicative of acceleration
in black hole disk dynamos. One more time, the emission spectra of
high-energy gamma rays could provide complementary information in this
context. Conspicuously, the diffuse neutrino intensity from RIAFs of
low luminosity active galactic nuclei can be compatible with the
observed IceCube data~\cite{Kimura:2014jba}.

The landmark detection of the gravitational-wave source
GW150914~\cite{Abbott:2016blz} opens another door for testing the 
Blandford-Znajek mechanism.  The GW150914 waveform indicates that the
source of the gravitational waves is the coalescence of two black
holes of rest-frame masses $36^{+5}_{-4} M_\odot$ and $29^{+4}_{-4}
M_\odot$, at luminosity distance $410^{+160}_{-180}~{\rm Mpc}$
(i.e. redshift $0.09^{+0.03}_{-0.04}$). The mass and spin
parameter of the newly formed black hole are 
\begin{equation}
M = 62^{+ 4}_{-4}
M_\odot  \quad
 {\rm and} \quad 
a = 0.67^{+0.05}_{-0.07} \,,
\label{LIGO-M}
\end{equation}
 and the source is localized
to a sky area of $600~{\rm deg}^2$.  

The maximum magnetic field
strength that could be steadily confined by the newly formed black hole can
be estimated assuming that there is rough equipartition of the magnetic
field energy density $B^2/(8\pi)$ with the radiation energy density,
and consequently with the accretion energy density. The energy density
of the radiation field at the Eddington limit is found to be
\begin{equation}
\rho_{\rm Edd} \sim \frac{L_{\rm Edd}}{4 \pi r_g^2 c} \sim 2 \times 10^{14}~{\rm  erg} \, {\rm cm^{-3}} \, ,
\end{equation}
and so the equipartition
field strength is 
\begin{equation}
B_{\rm Edd}  \sim \left( \frac{2 L_{\rm Edd}}{r_g^2 c} \right)^{1/2}
\sim 8 \times 10^7~{\rm G}  \,  ,   
\label{LIGO-B}
\end{equation}
Substituting (\ref{LIGO-M}) and (\ref{LIGO-B}) into (\ref{gamma-cr})
assuming $\zeta \sim r_g$ for protons we obtain $\gamma_{\rm max}^{\rm
  acc} \sim 10^8$. This is comparable to $\gamma_{\rm max}^{\rm rad}
\sim 10^{7.7}$. Collisions of the ultra-relativistic protons with the
ambient gas and photon backgrounds surrounding the black hole would
produce secondary neutrinos directly chasing the gravitational
waves~\cite{Moharana:2016xkz}. The IceCube and ANTARES collaborations
conducted a combined search for neutrinos candidates in both temporal
and spatial coincidence with GW150914, using a time window of $\pm
500~{\rm s}$ around the gravitational wave
transient~\cite{Adrian-Martinez:2016xgn}. No neutrino candidates have
been observed.

The remarkable power of GW150914 is open to more speculation.  The
total energy radiated in gravitational waves during GW150914 is found
to be~\cite{Abbott:2016blz} 
\begin{equation}
\varepsilon_{\rm GW} = (3.0 \pm 0.5) \ M_\odot \ c^2 \sim 5.4 \times
10^{54}~{\rm erg} \,, 
\end{equation}
with a peak wave luminosity of 
\begin{equation}
L_{\rm GW} = 3.6^{+0.5}_{-0.4} \times 10^{56}~{\rm erg} \, {\rm
  s}^{-1} \, . 
\end{equation}
Note that at the peak the luminosity in gravitational waves is more
that 15 orders of magnitude above the Eddington limit. Equipartition
arguments seem to indicate that rotationally-powered super-Eddington
magnetic fields of $B_{\rm sEdd} \sim 10^{11}~{\rm G}$ might have
existed for a limited period of time $\sim 50~{\rm
  ms}$~\cite{Kotera:2016dmp}. However, for such a large magnetic
field strength, the maximum proton energy will be limited by radiative losses;
namely,
\begin{eqnarray}
\gamma_{\rm max}^{\rm acc} & \sim & 10^8 \ \frac{Z}{A} \ \frac{B_{\rm sEdd}}{B_{\rm
    Edd}} \ \left(\frac{\zeta}{r_h} \right)^2 \, , \nonumber \\
\gamma_{\rm max}^{\rm rad} & \sim &  10^8 \ \left(\frac{a}{Z}
\right) \
\left(\frac{B_{\rm sEdd}}{B_{\rm Edd}} \right)^{1/4} \
\left(\frac{r_c}{r_h} \right)^{1/2} \  \left(\frac{\zeta}{r_h}
\right)^{1/4} \, .
\end{eqnarray}
Now, assuming  $\zeta  \sim r_c \sim 150~r_g$, we would expect to see a few protons reaching Lorentz factors of
up to $\sim 10^{10.3}$, in a time scale $\zeta/c \sim 50~{\rm ms} $.  Should
nature be so cooperative, since the proton spectrum is expected to be
sharply peaked at nearly the maximum energy, the GW150914 neutrino
emission would have been predominantly at ${\rm EeV}$ energies. Thus, the
search for temporal and spatial coincidence of these onliest prompt
EeV neutrinos with binary black hole mergers could provide direct
evidence
for cosmic ray acceleration to extreme energies in a one-shot boost.\\

In summary, the timing and approximate positional coincidences of one
of the IceCube neutrino events with the observed photon flaring in
X-rays at the Galactic center by Chandra represents the ``first light'' in the
nascent field of multiwavelength--multimessenger
astronomy~\cite{Bai:2014kba}. Herein we have put forward a possible
explanation for such intriguing association. Future X-ray observations
confronted with high-energy neutrino data will test the IceCube-Chandra
connection, providing the final verdict for the ideas discussed in
this paper.

\acknowledgments{LAA has been supported by the U.S. National Science
  Foundation (NSF) CAREER Award PHY1053663 and by the National
  Aeronautics and Space Administration (NASA) Grant No. NNX13AH52G; he
  thanks the Center for Cosmology and Particle Physics at New York
  University for its hospitality. Any opinions, findings, and
  conclusions or recommendations expressed in this material are those
  of the author and do not necessarily reflect the views of  NSF or
  NASA.}


\end{document}